\documentclass[letterpaper, 10 pt, conference]{ieeeconf}

\IEEEoverridecommandlockouts                              % This command is only
 \pdfoutput=1
\usepackage{amsmath, amssymb}
\usepackage{amsfonts}
\usepackage{graphicx}
\usepackage{color}
\usepackage{enumerate}
\usepackage{caption}
\usepackage[lofdepth,lotdepth]{subfig}
\graphicspath{{figures/}}

\newtheorem{rem}{Remark}

\newcommand{\thetah}{\hat{\theta}}
\newcommand{\omegah}{\hat{\omega}}
\newcommand{\wh}{\hat{\rm{w}}}
\newcommand{\rw}{\rm{w}}
\newcommand{\gb}{\bar{g}}
\newcommand{\yh}{\hat{y}}
\newcommand{\Hh}{\hat{H}}
\newcommand{\Xh}{\hat{X}}
\newcommand{\expec}[1]{ E \left\{ #1 \right\} }
\title{The Price of Synchrony: \\ Resistive Losses due to Phase Synchronization in Power Networks }
\author{Bassam Bamieh and Dennice F. Gayme %
\thanks{B. Bamieh is with the Department of Mechanical Engineering at the University of California at Santa Barbara, Santa Barbara, CA, USA, 93106.
        {\tt (bamieh@engineering.ucsb.edu)}
\newline D. F. Gayme is with the Department of Mechanical Engineering at the Johns Hopkins University, Baltimore, MD, USA, 21218. {\tt (dennice@jhu.edu)}. This work is partially supported by  AFOSR grant  FA9550-10-1-0143.}%
        }

\begin{document}

\maketitle
\thispagestyle{empty}
\pagestyle{empty}
\begin{abstract}

We investigate the total resistive losses incurred in returning a power network of identical generators to a synchronous state following a transient stability event or in maintaining this state in the presence of persistent stochastic disturbances. We formulate this cost as the input-output
$H^2$ norm of a linear dynamical system with distributed disturbances. We derive an expression for the total resistive
 losses that scales with the size of the network as well as properties of the generators and power lines, but is independent of the network topology. This topologically invariant scaling of what we term the price of synchrony is in contrast to typical power system stability notions like rate of convergence or the region of attraction for rotor-angle stability.
Our result indicates that highly connected power networks, whilst desirable for higher phase synchrony, do not offer an advantage
in terms of the total resistive power losses needed to achieve this synchrony. Furthermore, if power flow is the mechanism used to achieve
synchrony in highly-distributed-generation networks, the cost increases unboundedly with the number of generators.

 \iffalse(i.e. the amount of feedback required to reach consensus), which strongly depend on the underlying network graph \cite{Florian,Bergen and Hill??}.  \textbf{Also might want to discuss how it relates to the ability of the network to sync}\fi
\end{abstract}

\section{Introduction}
\label{sec:intro}
Many factors such as increased demand, renewable energy mandates \cite{doereportmandates} and further deregulation of the electric power industry \cite{windBook,Carrasco2006,EPG} are driving changes to the electric power grid. The new grid will have to deal with higher levels of uncertainty from renewable energy sources and changing load patterns as well as increasingly distributed energy generation.
These changes are likely to make it more prone to stability issues. In particular, they have the potential to create problems associated with rotor-angle stability, which is the ability of the power grid to recover synchrony after a disturbance \cite{Kundar2004}. Synchrony refers to the condition when both the frequency and phase of all generators within a particular power network are aligned. Loss of synchrony can lead to load shedding or even black-outs.\iffalse
\textcolor{red}{(should we say in addition ``referred to as synchronism in ref [5]''? we use the term synchrony, and i'd like to keep it. is
synchronism more standard in that literature? it sounds linguistically awkward) }\fi

\iffalse May want to distinguish between phase locking, which is frequency synchronization (i.e. a constant angle difference) and phase synchronization (i.e. when angles are equal).\fi
A special case of rotor-angle stability is the so-called transient stability problem, which is associated with large angle disturbances due to events such as generator, power line or other component failures or other abrupt changes that can be caused, for example, by intermittent renewable energy sources. There is a large body of power system transient stability literature, see e.g. \cite{Varaiya1985, Alberto2001}.  Most of this literature focuses on the existence of Lyapunov like energy functions \cite{Pecora1998, Nara_1984} and their use in determining a region of attraction type criteria for a particular synchronous state or set of states, see e.g. \cite{Chiang1988,Silva2005}.

%	Synchronism in power systems is analogous to the problem of a set of coupled dynamical systems converging to a common state. This is the so-called consensus problem, which has been extensively studied in the controls community \cite{Reza2007}.
%	{\color{red}relation to consensus problems?}

A recent research trend has been to draw connections between problems in distributed system control and power network stability.
This literature is vast, but  we note in particular a series of works that use a set of coupled Kuramoto oscillators to represent the power network~\cite{Dorfler2010,Dorfler2011}.  This non-uniform Kuramoto oscillator modeling framework uses a first order approximation of the network reduced classical power system model to provide network parameter dependent analytical conditions for frequency and phase synchronization~\cite{Dorfler2011}. Similar first order models have been employed to investigate how power flow scheduling and adding more power lines (i.e., increasing graph connectivity) affects the rate of convergence in a power network~\cite{Tang2011}.

In the present paper, we examine  the connections between distributed control systems and power network stability in a different
context. We do not study stability, but rather assume that the network will return to a synchronized state after disturbances. Instead we focus on the cost of keeping the network in synchrony, i.e. how much real power is required to drive the system to a stable, synchronous operating condition. Lack of synchrony leads to circulating currents \cite{vonMeier2006} passing between generators whose angles are out of phase. This flow of current leads to resistive power losses over the power lines due to their non-zero line resistances. This loss is generally considered relatively small compared to the total real power flow in a typical power network. It is however unclear whether these losses will be small in  power grids of the future, which are expected to have highly distributed generation,
and consequently many more (though typically smaller power capacity) generators than today's grid.

\iffalse
In the present paper, we take these connections between distributed control systems and power network stability in a  different
direction. We are not concerned with network stability per say, we actually assume that the network indeed has transient stability
and will return to a synchronized state after disturbances. What we study is the actual cost of keeping the network in synchrony.
AC power networks presently use the flow of reactive power back and forth between generators to keep them in synchrony. There is
a consequent resistive power loss in transmission between generators due to non-zero line resistances. This loss is
considered relatively small compared to the total real power flow in current networks. It is however unclear whether these losses
would be small in a future network with highly distributed generation on the order of say millions of generators.\fi

The problem we analyze is that of a large network of many identical generators. We consider several scenarios such as the power network
encountering disturbance (transient stability) events, or being subjected to persistent stochastic disturbances where the system is
continuously correcting for these disturbances. In both of these scenarios, we quantify the total power lost due to non-zero line
resistances. Our main result, in equation \eqref{eqn:mainresult}, shows that this power loss scales  with the product of
the {\em network size} and the {\em ratio of line resistances to their reactances} (which is assumed to be the same for all links).
The latter quantity is normally assumed to be rather small. However,
our result shows that even though that ratio is small,  when the number of generators becomes very large, then so will the total resistive losses. Furthermore, and perhaps
more surprisingly, these losses are independent of the network topology, i.e. highly connected networks (which would have highly
coherent, or close to identical, phases) and loosely connected networks (which would have a higher degree of phase incoherence) incur the same
resistive power losses in recovering synchrony.

Our results indicate that the cost of maintaining synchrony using power flows is
essentially a function of the number of generators in the network and not of its topology.
We should point out that this conclusion is not inconsistent with other results on
power system stability and performance measures. For example,  interaction strength and network topology play important roles in determining whether a system can synchronize \cite{Pecora1998,Dorfler2010,Dorfler2011,Bamieh2012} and the rate of convergence or damping of a power system is directly related to the network connectivity \cite{Tang2011}.
The numerical examples we present in this paper confirm that while  losses are independent of the network connectivity,
 a highly connected network will stabilize more quickly and with less oscillatory behavior.

 One measure that quantifies the lack of synchrony between network elements is that of ``coherence''~\cite{Bamieh2012}.
 As expected, highly connected networks are more coherent than loosely connected ones. We refer the reader to~\cite{Bamieh2012} for
 asymptotic scalings of lattice-type networks and to~\cite{patterson2011network} for fractal networks, and note that these scalings
 are somewhat different than those that quantify network damping.
 One intuitive explanation for the cost of synchrony being
 independent of network topology is as follows. Comparing a highly connected network with a loosely connected one, we expect
 the former to have much more phase coherence than the latter. Consequently the power flows per link in a highly connected network
 are relatively small, but there are many more links than in the loosely connected network. Thus in the aggregate, the total power losses
 of the two networks are the same. One should keep in mind however, that a less coherent network is more likely to have transient
 stability problems, exit the region of attraction, etc. The issues  of stability and the cost of synchrony are two different issues.

The remainder of this paper is organized as follows.  Section \ref{sec:ProblemSetup} introduces the classical power system model and defines the notation.  In Section \ref{sec:performance} we quantify total resistive line losses as an input-output  $H^2$ norm of a
linear dynamical system.
This framework is then used to derive an algebraic expression for the resistive line losses in terms of the parameters of the admittance matrix, generator damping and the size of the network. Various interpretations of the $H^2$ norm are then presented together with
the corresponding operating scenarios for a power network . The results of the analysis are demonstrated for two different network topologies in Section \ref{sec:examples}. The paper concludes and suggests directions for future work in Section \ref{sec:conclusions}.

\iffalse
The amount of control or actuation required for a system to attain a desired condition is known to be an important consideration in control design and analysis. In a consensus setting this question is related to the control effort required for a distributed system to perform a coordinated task {\color{red}\cite{}}. In the power system context this question is related to how much power is needed to drive the system to a stable, synchronous operating condition. This quantity can be measured by looking at the resistive losses based on circulating currents caused by lack of synchrony. The present work characterizes this power loss (cost) in a system of identical coupled oscillators subjected to stochastic disturbances. The results show that this cost scales with the size of the network, generator damping and ratio of power line resistance to reactance, but does not depend on the network topology.
\fi

\section{Problem Formulation}
\label{sec:ProblemSetup}
 Consider a network of $N$ buses (nodes) and $\mathcal{E}$ edges. At each node $i=1,\dots, N$ there is a generator ${G}_i$, with inertia constant $M_i$, damping $\beta_i$, voltage magnitude $V_i$ and angle $\theta_i$.  Using the classic machine model, which assumes a fixed voltage behind a reactance, the dynamics of the $i^{th}$ generator is given by~\cite{Pai1981}
\begin{equation}
M_i \ddot{\theta}_i+\beta\dot\theta_i=P_{m,i}-P_{e,i}\;\;\;\;\;\;\forall\;
i=1,2...n. \label{eqn:swing}
\end{equation}
Here, $P_{e,i}$ is the electrical power flowing out of the $i^{th}$ generator, and $P_{m,i}$ is the mechanical power input from the corresponding turbine.

\begin{figure}
\centering
\includegraphics[width=0.45\textwidth,clip]{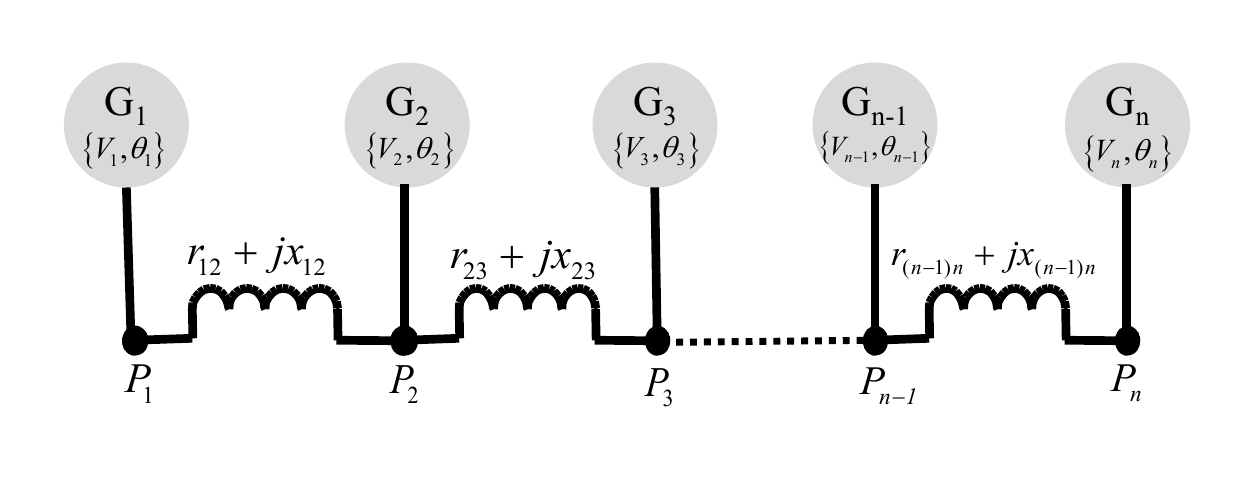}
\caption{A depiction of a linear network of $n$ generator nodes (buses) $G_i$ for $i=1\dots,n$. }\label{fig:gen_string}

\end{figure}

Define the admittance over edge $\mathcal{E}_{ij}$ connecting nodes $i$ and $j$ as $y_{ij}=g_{ij}-\textbf{j}b_{ij}$, where $g_{ij}$ and $b_{ij}$ are respectively the conductance and susceptance of the line defined by the edge $\mathcal{E}_{ij}$. Then, the electrical power injection at node $i$ is
\begin{multline}
\label{eqn:pwrflow}
P_{e,i}~=~ g_{ii} |V_i|^2+ \sum_{k\sim i}g_{ik}|V_i|\,|V_k|\cos(\theta_i-\theta_{k})\\
\quad\quad+\sum_{k\sim i}b_{ik}|V_i|\,|V_k|\sin(\theta_i-\theta_{k}),
\end{multline}
where $k\sim i$ denotes an edge $\mathcal{E}_{ik}$ and $g_{ii}$ is the self admittance of the generator. In what follows, we use a simplified model in which all generators are assumed to have the same self admittance, which we denote $\gb =g_{ii}$ for all $i=1,\dots,n$. Finally, applying the standard linear power flow assumptions used in transient stability analysis (conductances are negligible, angle differences are small and voltages are constant with unit magnitude) to \eqref{eqn:pwrflow} yields
\begin{equation}
\label{eqn:lin-pwrflow}
P_{e,i}\approx \sum_{k\in\mathcal{N}}b_{ik}\left[\theta_i-\theta_{k}\right].
\end{equation}
Substituting this into \eqref{eqn:swing} leads to
\begin{equation}
\label{eqn:second_order_full}
M_i \ddot{\theta}_i+\beta\dot\theta_i\approx- \sum_{k\in\mathcal{N}}b_{ik}\left[\theta_i-\theta_{k}\right]+P_{m,i}.
\end{equation}

In order to simplify the notation we define the entries of the admittance matrix $Y \in \mathbb{C}^{n \times n}$ as
\begin{align*}
Y_{ij} := \begin{cases} \gb + \displaystyle\sum_{j \sim i} y_{ij}, & \text{ if } i=j,\\
- y_{ij}, & \text{ if } i \neq j \text { and } j \sim i, \mbox{ (i.e., } \mathcal{E}_{ij} \in \mathcal{E}), \\
0 & \text{ otherwise}.\end{cases}
\end{align*}
Then, we partition this admittance matrix into the real (resistive) and imaginary (reactive) parts and define
\[
	Y~=~ \text{Re}\{Y\}~+~\textbf{j}~\text{Im}\{Y\} ~=:~ (L_G+\gb I) ~+~ \textbf{j}~L_B.
\]
By this construction, $L_G$ and $L_B$ retain the symmetry of $Y$, and they have as a common eigenvector
 the vector $\bold{1}$ with components all equal to 1, i.e.
\[
	L_B \bold{1} ~=~L_G \bold{1} ~=~ 0.
\]
It is a well-known result that if the graphs underlying the system represented by $L_B$ and $L_G$, which are analogous to weighted graph Laplacians, are connected (i.e. any two nodes are connected by a path of edges), then the remaining eigenvalues of $L_B$ and $L_G$  are all positive.

An assumption we will invoke later is that all the eigenvectors of $L_B$ and $L_G$ are shared ($u$ is an eigenvector
of $L_B$ iff it is an eigenvector of $L_B$). An important consequence of this assumption is that the two Laplacians
are simultaneously diagonalizable by the same orthogonal transformation $U$, i.e.
\begin{equation}
   \label{eqn:diagonalization}
	U^* L_B U ~=~ \Lambda_B , ~~~~
	U^* L_G U ~=~ \Lambda_G,
\end{equation}
where $\Lambda_B$ and $\Lambda_G$ are diagonal matrices with the respective eigenvalues of $L_B$ and $L_G$ as diagonal entries. One setting in which this assumption holds is when the ratios of each connection's conductance to
susceptance (equivalently the ratios of resistances to reactances) are all equal, i.e.
\[
	\frac{g_{ik}}{b_{ik}} ~=~ \frac{r_{ik}}{x_{ik}} ~=~ \alpha,
\]
which is independent of the link indices $(i,k)$. It then follows that
\begin{equation}
   \label{eqn:lglb}
	L_G ~=~ \alpha L_B  ,
\end{equation}
which implies that $L_G$ and $L_B$ have the same eigenvectors.

Finally, we rewrite equation
 \eqref{eqn:second_order_full} in state space form
\begin{eqnarray}
\label{eqn:state_space}
\frac{d}{dt} \begin{bmatrix} \theta (t) \\  \omega (t) \end{bmatrix} & = &
\begin{bmatrix}
0 & I\\
-L_{B} &-\beta I
\end{bmatrix}
 	\begin{bmatrix} \theta (t) \\  \omega (t) \end{bmatrix} 	~+~
 \begin{bmatrix}
0 \\ I
\end{bmatrix} \rm{w},			\\
 & =:& A \psi ~+~ B \rm{w} ,			\nonumber
\end{eqnarray}
where we have assumed that the constant $M_i=1$ for all $i$ and that mechanical power input $P_{m,i}$ is a constant that can be captured as a part of the input $\rm{w}.$

\begin{rem}
Based on the classical machine model $P_{m,i}$ is essentially the real power input at the bus. Thus one can consider $P_{m,i}$  as a system input. Alternatively, the absence of $P_{m,i}$ can be justified as follows.  For a system with no loads the equilibrium point of the system is at $P_{e,i}=P_{m,i}$.  If \eqref{eqn:swing} defines a linear system (as assumed here) a coordinate transform can be applied to obtain a new system with shifted angles $\theta^s$ corresponding to the transformed system with the equilibrium point $P_{m,i}=0$.
By abuse of notation we denote the shifted coordinates as $\theta$ in the state space representation \eqref{eqn:state_space}.
\end{rem}
\begin{rem}
The network model \eqref{eqn:state_space} assumes a system with no loads. Such a model may arise from network reduction leading to a system of internal generator buses with loads approximated as impedances that have been absorbed into the network. The model could easily be extended to include static loads through a system of differential algebraic equations similar to those derived in \cite{Tang2011,Hill2006}.
\end{rem}

\section{System Performance}
\label{sec:performance}
Several performance metrics can be used to quantify the relative stability for the system \eqref{eqn:state_space}. A few common metrics are the ability to synchronize, the degree of synchronization that is achievable, the time to synchronize, or the control effort required to obtain the desired system state. In the distributed systems setting it is also common to evaluate these metrics with respect to various control strategies. For example, in evaluating the performance of a distributed system with local versus global control strategies. In this section we focus on the control actuation required to drive a system to the synchronous state.  This control effort is measured through the real (resistive) power loss over each line.  These losses are associated with circulating currents that arise from the angle differences between generators, i.e. disturbances \cite{vonMeier2006}.

The power flow over an edge $\mathcal{E}_{ij}$ is
\begin{displaymath}
P_{ij}+P_{ji}=V_i\left(V_i-V_j\right)^*y_{ij}+V_j\left(V_j-V_i\right)^*y_{ji},
\end{displaymath}
where $^*$ denotes the complex conjugate. The resistive power loss over $\mathcal{E}_{ij}$ can therefore be defined as
\begin{equation}
P^{loss}_{ik} :=g_{ik}\left|V_i-V_k\right|^2.
\end{equation}
Using a small angle approximation and standard trigonometric identities this can be approximated as
\begin{equation}
\label{eqn:losses}
\tilde{P}^{loss}_{ik} = g_{ik}\left|\theta_i-\theta_k\right|^2.
\end{equation}
We are interested in the sum  total of the resistive losses over all links in the network, which is given by
\begin{equation}
\label{eqn:tot_loss}
\tilde{\mathbf{P}}_{loss}= \sum_{i\sim k}g_{ik}\left|\theta_i-\theta_k\right|^2.
\end{equation}
This last quantity can be expressed in vector form  as
\[
	\tilde{\mathbf{P}}_{loss}=y^*y
\]
where
the vector signal $y$ is an output of the linearized swing equations  \eqref{eqn:state_space}
\begin{equation}
\label{eqn:output}
y~=~ C\psi ~=:~ \begin{bmatrix}
C_1 & 0
\end{bmatrix}	\begin{bmatrix} \theta \\  \omega  \end{bmatrix} , ~~
\mbox{with~} C_1^* C_1:=L_G .
\end{equation}
A simple choice for $C_1$ would be to take $C_1 = L_G^{\frac{1}{2}}$ (possible since $L_G$ is
positive semi-definite),
which is what we assume from now on.

We now calculate the $H^2$ norm from disturbance $\rm{w}$ to the performance output $y$
 of the system  \eqref{eqn:state_space} and \eqref{eqn:output}. The square of the
  $H^2$ norm has several standard interpretations including
 {\em (a)} The variance of the output $y$ when the input $\rm{w}$ is a unit variance white stochastic process, {\em (b)}
  The total time integral
 of the variance of $y$ when the
 initial condition is a random variable with correlation matrix $BB^*$, and {\em (c)} The total sum of time integrals of output response
 powers given an impulse as a disturbance input at each generator.

 We first perform the $H^2$ norm calculation and derive a formula in terms of the system's parameters, and then investigate the
 implications of the three different interpretations for this particular system of swing equations.

 \subsection*{$H^2$ Norm Calculation}

 For ease of reference we rewrite the system equations \eqref{eqn:state_space} and \eqref{eqn:output} here
 \begin{equation}
 \label{eqn:systemall}
 	\begin{array}{rcl}
		\frac{d}{dt} \begin{bmatrix} \theta  \\  \omega  \end{bmatrix}
			& = &
					\begin{bmatrix}
					0 & I\\
					-L_{B} &-\beta I
					\end{bmatrix}
		 	\begin{bmatrix} \theta  \\  \omega  \end{bmatrix} 	
		~+~
			 \begin{bmatrix}
			0 \\ I
			\end{bmatrix} \rm{w} 			\\
		y & = & \begin{bmatrix}
				L_G^{\frac{1}{2}} & 0
				\end{bmatrix}	\begin{bmatrix} \theta \\  \omega  \end{bmatrix}
	\end{array} .
 \end{equation}
 We will denote the input-output mapping of this system by $H$.
 This system has all eigenvalues strictly in the left half of the complex plane with the exception of a single zero eigenvalue of $L_B.$
 It is, however easy to show that this  unstable mode is unobservable from the output $y$ due to the fact that
 $L_G=C_1^* C_1$ shares
 this eigenvalue and its corresponding eigenvector (see Appendix for the full argument).
 It then follows that the input-output transfer function from $\rm{w}$ to $y$ is indeed
 stable and has finite $H^2$ norm.

The easiest method to calculate the $H^2$ norm of the  \eqref{eqn:systemall} is using a spectral decomposition
of $L_B$. Consider the state transformation
\[
	\begin{bmatrix} \theta \\ \omega  \end{bmatrix}
	~=:~
	\begin{bmatrix} U & 0 \\ 0 & U \end{bmatrix} \begin{bmatrix} \thetah \\ \omegah  \end{bmatrix} ,
\]
where $U$ is the matrix in \eqref{eqn:diagonalization} diagonalizing $L_B$ and $L_G$. The system in the new state variables
becomes
\[
 	\begin{array}{rcl}
		\frac{d}{dt} \begin{bmatrix} \thetah  \\  \omegah  \end{bmatrix}
			& = &
					\begin{bmatrix}
					0 & I\\
					-\Lambda_{B} &-\beta I
					\end{bmatrix}
		 	\begin{bmatrix} \theta  \\  \omega  \end{bmatrix} 	
		~+~
			 \begin{bmatrix}
			0 \\ U^*
			\end{bmatrix} \rm{w} 			\\
		y & = & \begin{bmatrix}
				L_G^{\frac{1}{2}} U & 0
				\end{bmatrix}	\begin{bmatrix} \theta \\  \omega  \end{bmatrix}
	\end{array} .
\]
Since multiplying by orthogonal matrices does not change the $H^2$ norm, we can multiply the input
by $U$ and and the output by
$U^*$ (i.e. define $\wh = U^*\rm{w}$ and $\yh = U^* y$) to obtain an equivalent system (that has the same $H^2$ norm)
\[
 	\begin{array}{rcl}
		\frac{d}{dt} \begin{bmatrix} \thetah  \\  \omegah  \end{bmatrix}
			& = &
					\begin{bmatrix}
					0 & I\\
					-\Lambda_{B} &-\beta I
					\end{bmatrix}
		 	\begin{bmatrix} \theta  \\  \omega  \end{bmatrix} 	
		~+~
			 \begin{bmatrix}
			0 \\ I
			\end{bmatrix} \wh 			\\
		\yh & = & \begin{bmatrix}
				\Lambda_G^{\frac{1}{2}}  & 0
				\end{bmatrix}	\begin{bmatrix} \theta \\  \omega  \end{bmatrix}
	\end{array} .
\]
We will denote the input-output mapping of this system by $\hat{H}$.
Since $\Lambda_B$ and $\Lambda_G$ are diagonal, this represents $N$ decoupled systems
\[
 	\begin{array}{rcl}
		\frac{d}{dt} \begin{bmatrix} \thetah_n  \\  \omegah_n  \end{bmatrix}
			& = &
					\begin{bmatrix}
					0 & 1\\
					-\lambda^{B}_n &-\beta
					\end{bmatrix}
		 	\begin{bmatrix} \theta_n  \\  \omega_n  \end{bmatrix} 	
		~+~
			 \begin{bmatrix}
			0 \\ 1
			\end{bmatrix} \wh_n			\\
		\yh_n & = & \begin{bmatrix}
				\sqrt{\lambda^G_n}  & 0
				\end{bmatrix}	\begin{bmatrix} \theta_n \\  \omega_n  \end{bmatrix}
	\end{array},
\]
where $n =1,\ldots, N$ are the indices of eigenvalues $\lambda_n^B$ and $\lambda_n^G$, which correspond to $L_B$ and $L_G$
respectively. Denote the input-output mapping of each decoupled subsystem by $\hat{H}_n$. We can then
write
\[
	\Hh ~=~ \mbox{diag} \left( \Hh_1, \ldots, \Hh_N \right).
\]
The square of the  $H^2$ norm of the system \eqref{eqn:systemall} is thus the sum of the squares of the $H^2$ norms of
all the decoupled subsystems
\[
	\left\| H \right\|_2^2 ~=~ \|\Hh \|_2^2 ~=~ \sum_{n=1}^N \|\Hh_n\|_2^2.
\]
The $H^2$ norm of each of the individual subsystems can  now be  easily calculated as follows.
For $n=1$, the corresponding eigenvalues are $\lambda_1^B=\lambda_1^G=0$, and we have a
completely unobservable system, thus $\|\Hh_1\|_2=0$. For $n\neq 1$, we solve the Lyapunov
equation for the observability Grammians $\Xh_n$
\begin{eqnarray*}
	\begin{bmatrix} 0 & -\lambda^B_n\\ 1 & -\beta   \end{bmatrix}
	\begin{bmatrix}  \Xh_{11} & \Xh_{0} \\  \Xh_{0}^* & \Xh_{22}    \end{bmatrix}
	&+&
	\begin{bmatrix}  \Xh_{11} & \Xh_{0} \\  \Xh_{0}^* & \Xh_{22}    \end{bmatrix}
	 \begin{bmatrix}  0 & 1\\  -\lambda^B_n & -\beta    \end{bmatrix}				\\
	& = &-
	\begin{bmatrix}  \lambda^G_n & 0  \\   0 & 0   \end{bmatrix},
\end{eqnarray*}
where for simplicity of notation we have dropped the subscript $n$ from the components of
of the Grammian $\Xh_n$. This matrix equation corresponds  three equations, of which only the
following two are relevant
\[
\begin{aligned}
	-\lambda^B_n \Xh_{0}^*-\Xh_0 \lambda^B_n   = & -\lambda^G_n
	&\Rightarrow&
	\rm{Re}(\Xh_0)  =  \frac{1}{2} \frac{\lambda_n^G}{\lambda_n^B}			\\
	\Xh_{0}+\Xh_{0}^* -2\beta \Xh_{22} = & 0
	&\Rightarrow&
	\Xh_{22}		= \frac{1}{\beta} \rm{Re} (\Xh_0).
\end{aligned}		
\]
Finally, since the $B$ matrix of each subsystem is $\left[ 0 ~ 1 \right]^T$, the $H^2$ norm (squared) of each
subsystem is just $\Xh_{22}$, and we therefore conclude that
\[
	\|\Hh_n \|_2^2 ~=~ \frac{1}{2\beta} \frac{\lambda^G_n}{\lambda^B_n} .
\]
The total $H^2$ norm (squared) of the overall system \eqref{eqn:systemall} is thus
\begin{eqnarray*}
	\|H \|_2^2 & = &  \frac{1}{2\beta} \sum_{n=2}^N \frac{\lambda^G_n}{\lambda^B_n}
			~=~  \frac{1}{2\beta} \sum_{n=2}^N \frac{\alpha \lambda^B_n}{\lambda^B_n}		\\
			& = & \frac{\alpha}{2\beta} (N-1) .
\end{eqnarray*}

In summary, we conclude that the total resistive losses in this network are
\begin{equation}
  \label{eqn:mainresult}
  \boxed{
	\|H \|_2^2 ~=~ \frac{1}{\beta} \frac{r}{x} ~(N-1) ,
	}
\end{equation}
where $\beta$ is a generator's self damping, $\frac{r}{x}$ is the ratio of a line's resistance to its reactance
(assumed equal for all lines), and $N$ is the number of generators in the network. Note that this expression
is independent of the network topology but depends on the number of generators in the network.

\begin{figure*}
  \centering
  \subfloat[][5 Node Line Graph]{
  \includegraphics[width = 0.425\textwidth,clip]{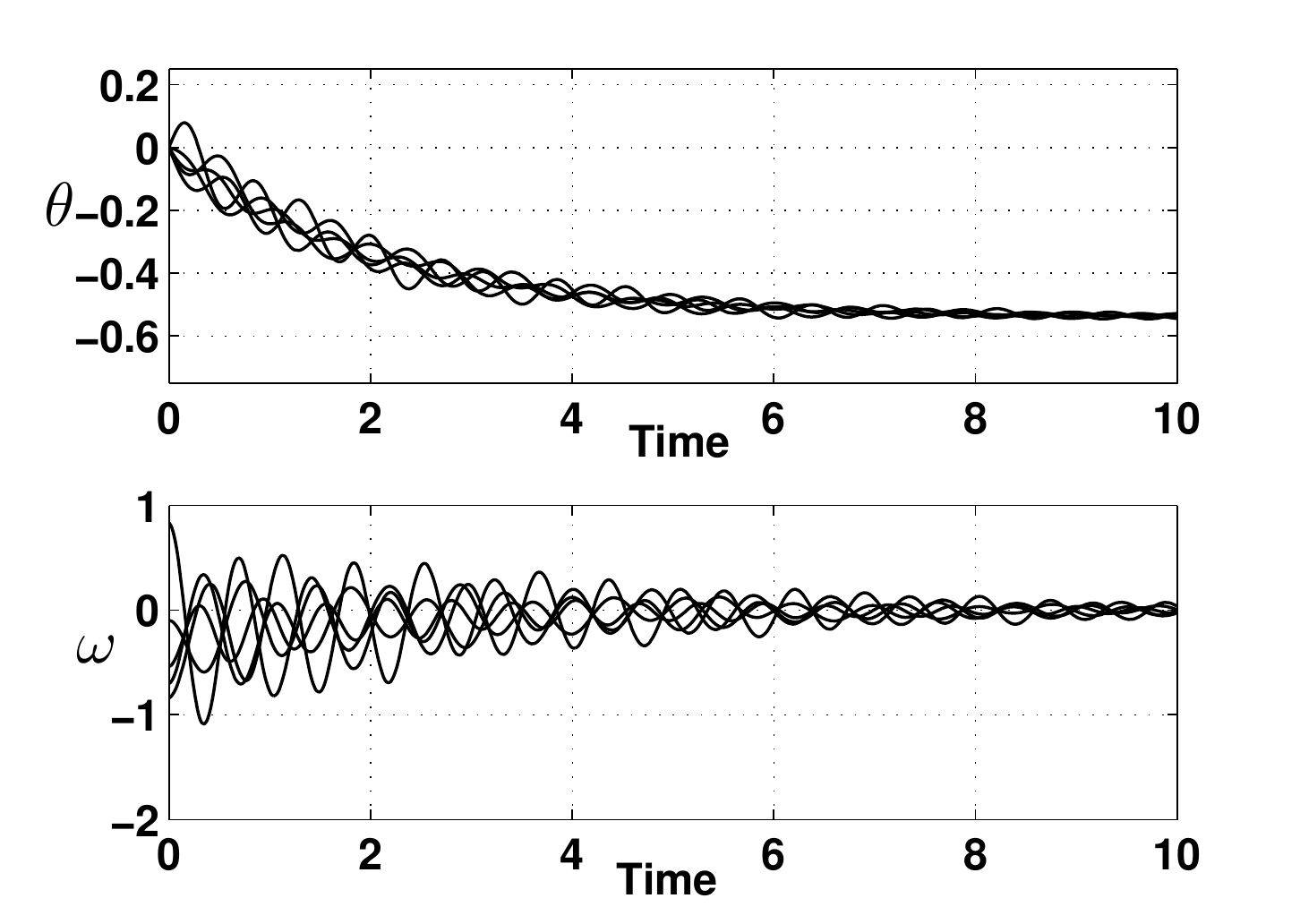}
  \label{fig:line-5}  }
  \subfloat[][Fully Connected 5 Node Graph] {
  \includegraphics[width = 0.425\textwidth,clip]{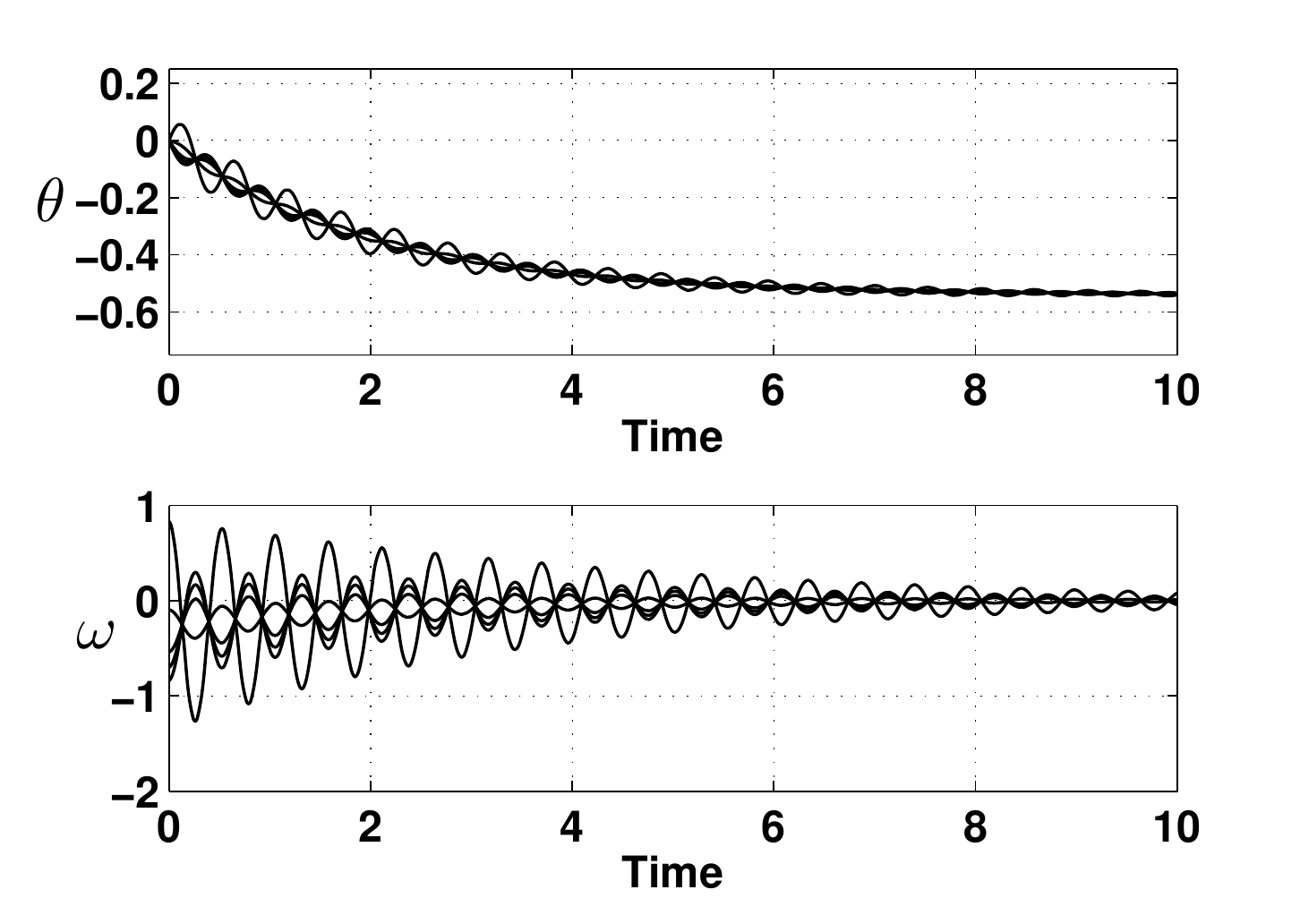}
  \label{fig:full_connect-5}}\\
   \subfloat[][20 Node Line Graph]{
  \includegraphics[width = 0.425\textwidth,clip]{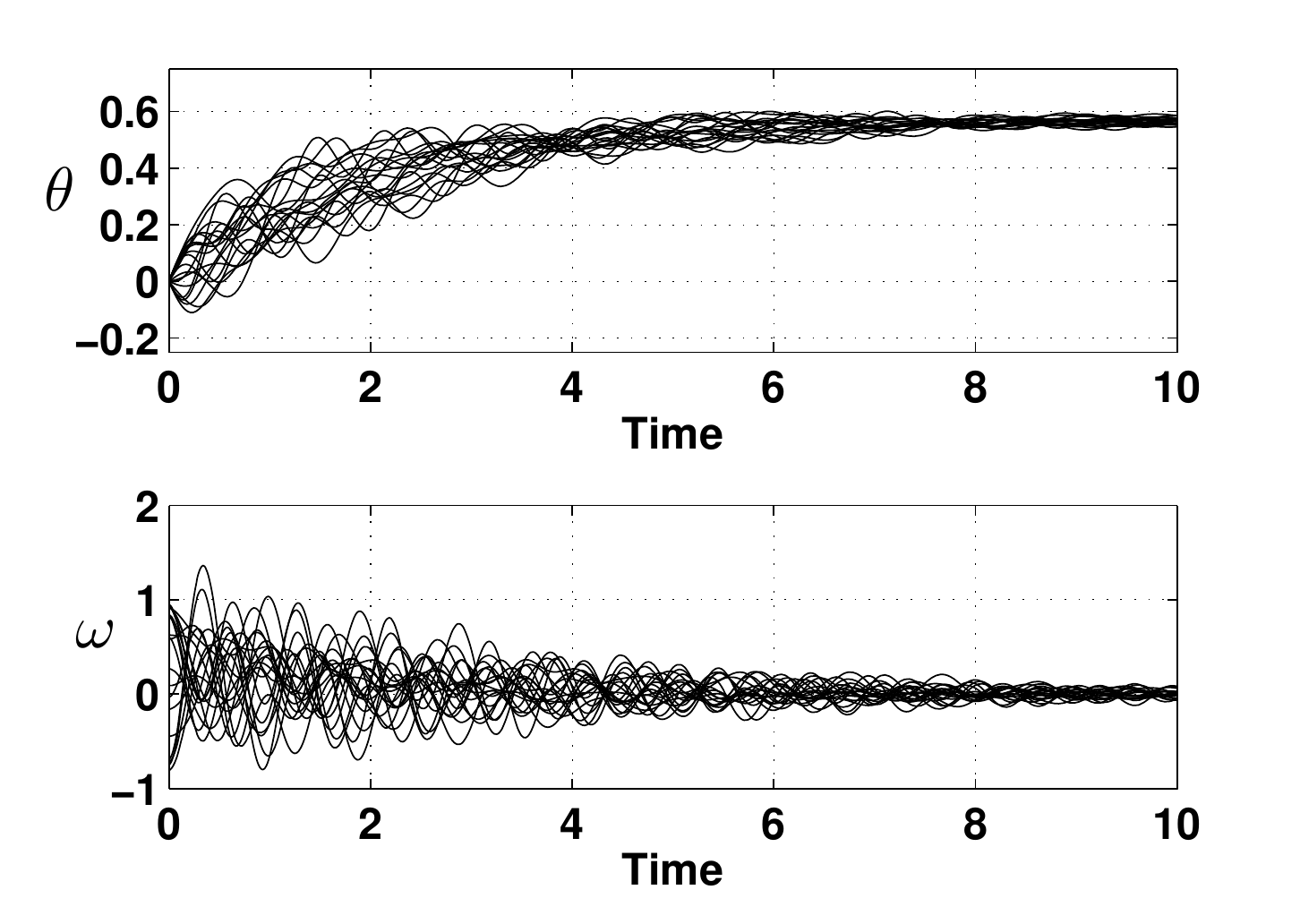}
  \label{fig:line-20}  }
  \subfloat[][Fully Connected 20 Node Graph] {
  \includegraphics[width = 0.425\textwidth,clip]{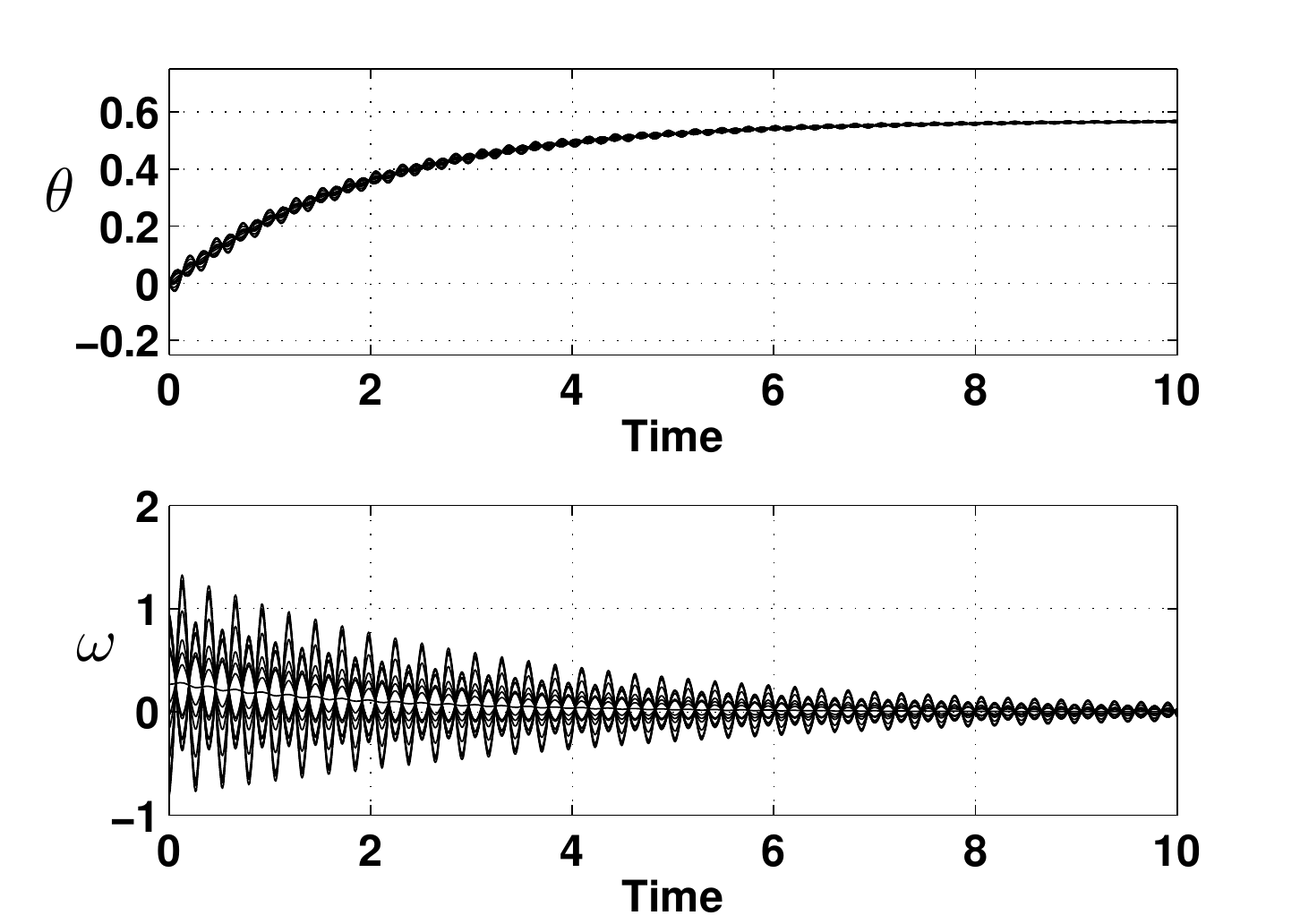}
  \label{fig:full_connect-20}}
    \caption{(a),(b) Simulation of  a $5$ bus system of identical oscillators with a random initial velocity disturbance, and zero
    		initial phase disturbance.   (c),(d) Simulation of a system with 20 identical oscillators under the same conditions as (a),(b).
		Figures (a),(c) are for linearly connected networks (with line graphs as in Figure \ref{fig:gen_string}),
		while Figures (b),(d) are for fully connected networks (complete graphs).
		Note how linear networks are much less coherent than complete networks (Figures (a),(c) versus (b),(d) respectively).
		Total resistive losses are however the same for the 5 node networks (a) and (b), and for the 20 node networks,(c) and (d),
		respectively, reflecting that resistive line losses are a function of the network's size rather than its topology. }
\label{fig:Sims}
\end{figure*}

\subsection*{$H^2$ Norm Interpretations for Swing Dynamics}

	In the previous subsection we calculated the $H^2$ norm of the linearized swing dynamics  \eqref{eqn:state_space}, together with the output equation  \eqref{eqn:output} based on a disturbance input (forcing) $\rm{w}$. In our formulation the square of the  Euclidean norm $y^*y$ of the output vector was defined to be equal to the total real power dissipated due to resistive losses in the network connections as the system synchronized. This synchronized state (the equilibrium) is maintained by circulating currents leading to power flowing back and forth between generator nodes.
	
	It is well known that the $H^2$ norm has at least three different interpretations. We recall these interpretations
	here in order to give three different physical scenarios in which equation \eqref{eqn:mainresult} quantifies the
	resistive losses due to the synchrony requirement. Denote by $H$ the following Linear Time Invariant (LTI) system
	\begin{eqnarray*}
		\dot{\psi}(t) & = &  A \psi(t) ~+~ B \rm{w}(t) 	\\
		y(t) & = & C \psi(t) ,
	\end{eqnarray*}
	The $H^2$ norm $\|H\|_2$  of the system  has the following three different interpretations.
	\begin{enumerate}[(a)]
		\item {\em Response to a white stochastic input.} When the input $\rm{w}$ is a white second order
				process with unit covariance (i.e. $\expec{ \rw(t) \rw^*(t)} = I$), the $H^2$ norm (squared) of the system
				is the steady-state  total variance of all the output components, i.e.
				\[
					\|H\|_2^2 ~=~ \lim_{t\rightarrow \infty}  \expec{y^*(t) y(t) } .
				\]
				
				For the swing dynamics \eqref{eqn:systemall} the  disturbance vector  $\rm$ can be thought of
				as persistent stochastic disturbances (forcing) at each generator. These disturbances, which are uncorrelated across
				generators, can be due to uncertainties in local generator conditions, sudden changes in load, or fault events. The variance of the output is exactly the expectation of the total power loss due to line resistances.
								
		\item {\em Response to a random initial condition.} When the input is turned off, but the initial condition is a random
				variable $x_o$ with correlation $\expec{x_o x_o^*} = BB^*$, then the $H^2$ norm is the time integral
				\[
					\|H\|_2^2 ~=~ \int_0^\infty  \expec{y^*(t) y(t) } ~dt
				\]
				of the resulting response $y$.
				
				The interpretation for  \eqref{eqn:systemall} is as follows. Since
				$BB^* = \begin{bmatrix} 0 & 0  \\ 0 & I   \end{bmatrix}$, the corresponding initial condition corresponds to
				each generator having a random initial velocity perturbation that is
				uncorrelated across generators and zero initial phase perturbation. Then $\|H\|_2^2$  quantifies the total (over all time) expected
				resistive power losses due to the system returning to a synchronized state.
		
		\item {\em Sum of responses to impulses at all inputs.} Let $e_i$ refer to the vector with all components zero
				expect for $1$ in the $i^{th}$ component. Consider $N$ experiments where in each,  the system is fed an impulse
				at the $i^{th}$ input channel, i.e. $\rm{w}_i(t) ~=~ e_i \delta(t)$. Denote the corresponding output by $y_i$. The
				$H^2$ norm (squared) is then the sum total of the $L^2$ norms of these outputs, i.e.
				\[
					\|H\|_2^2 ~=~ \sum_{i=1}^N \int_0^\infty y_i^*(t) y_i(t) ~dt .
				\]
				A stochastic version of this scenario corresponds to a system where the inputs $\rw_i$ can occur with equal probability. Under this assumption the $\|H\|_2^2$ becomes the expected total power loss given these inputs.
				
				The interpretation for  \eqref{eqn:systemall} is when each of the generators are subject to impulse force
				disturbances (since $\rm$ enters into the momentum equation of each generator), and $\|H\|_2^2$ is thus
				the total power loss over all time under such a scenario.

	\end{enumerate}

\section{Numerical Examples}
\label{sec:examples}

Consider two networks of identical generators, one whose underlying graph is a line, as in Figure \ref{fig:gen_string}, and one with a fully connected graph. In this section, we compare the behavior of systems with these two network topologies as the network size varies. All simulations use the following parameter values \cite{SauerBook}: $M=\frac{20}{2\pi f}$, $\beta=\frac{10}{2\pi f}$ with a frequency $f=60$ Hz. The admittance between connected generators is set to $y_{ij}=0.2+\mathbf{j}1.5$  and we assume $g_{ii}=:\gb=0$ for all $i=1,\dots,n.$

Figures \ref{fig:line-5} and \ref{fig:full_connect-5} respectively show the state trajectories of a 5 node system with a line graph and one with a fully connected graph given identical initial conditions. The initial conditions for each $\omega_i(0)$ were drawn from a uniform distribution $[-1,1],$ and $\theta(0)$ was set to zero. This corresponds to the $H^2$ norm interpretation (b) described in the previous section. For both the  line graph and the fully connected network the total losses are the same, with $\mathbf{P}^{loss}=1.2885$, as predicted by the main result in \eqref{eqn:mainresult}. However, the transient behavior shows that the system with the fully connected graph
has more ``coherent'' phases (i.e. they stay closer together).
This is most clearly visible in the top panels of the plots that depict the state $\theta$. Here, the small oscillations for the line graph system continue through the $10$ second interval shown in Figure \ref{fig:line-5} but have almost completely died out for the fully connected of Figure \ref{fig:full_connect-5}.

In order to evaluate the effect of increasing the network size we ran a similar test with two 20 bus systems with the same topological structure and initial conditions ($\omega_i(0)\in[-1,1]$ drawn from a uniform distribution for each $i$ and $\theta(0)=0$).  In this case, the losses increased to $\mathbf{P}^{loss}=5.9256$ but remained equal in both networks. The faster convergence to the synchronized state in fully connected system (compared to the line graph) is much more evident in the larger network, as shown in figures \ref{fig:line-20} and \ref{fig:full_connect-20}.

The simulation results indicate that a fully connected graph is much more coherent than a line graph. The resemblance of this
result to those about coherence in vehicular formations (platoons) and similar consensus-like
 network algorithms~\cite{Bamieh2012} is striking.  With coherence as a performance measure, the line graph is the worst
 such topology (amongst connected graphs), while the fully connected topology is the best.
However this additional coherence, while desirable for other reasons, will  not result in lower resistive losses.

\section{Conclusions}
\label{sec:conclusions}

We have considered  a power network model with distributed disturbances, and quantified the total resistive power losses incurred
due to the current flows needed to maintain phase synchrony. We have shown that these losses are independent of the network's
topology, but scale unboundedly with the number of generators. There are interesting implications for the design of future
highly-distributed-generation networks, which have potentially orders of magnitude more generator nodes that today's networks.
While a highly phase coherent (and therefore highly connected) network is desirable for many reasons, the cost of maintaining
this coherence depends only on the number of generators and not the network's connectivity. Since this cost grows unboundedly
with the number of generator nodes, then the current scheme of using power flows as the synchronization mechanism may not be
scalable to future networks. This is perhaps a further argument for the use of other control mechanisms, such as communication
links, for phase synchronization.

%
%Thus there is no benefit from a power loss perspective in designing a highly connected power system. However, a badly connected system will longer to settle back to a synchronized equilibrium state after a system disturbance. The additional oscillations and associated cyclic loading involved in resynchronizing the system may be detrimental to the life cycle generator.
%

\section*{Appendix}

\subsection*{Proof that  \eqref{eqn:systemall} is input-output stable}

It is well known that for any pair $(C,A)$, the observability of $(C,A)$ and $(C^*C,A)$ are equivalent.
The only unstable mode of the A-matrix in  \eqref{eqn:systemall} is at zero with corresponding eigenvector
$\begin{bmatrix} \bold{1} \\ 0  \end{bmatrix}$. This mode is unobservable from the output $y$ since by the
PBH test
\[
	\begin{bmatrix} -A \\ C^*C  \end{bmatrix}
	\begin{bmatrix} \bold{1} \\ 0  \end{bmatrix}
	~=~
	\begin{bmatrix}
		\begin{bmatrix} 0 & -I \\  L_{B} &\beta I  \end{bmatrix}
		 			\\
		\begin{bmatrix} L_G & 0 \\ 0 & 0 \end{bmatrix}
	\end{bmatrix}
	\begin{bmatrix} \bold{1} \\ 0  \end{bmatrix}	
	~=~
	\begin{bmatrix}0 \\ 0  \end{bmatrix}	.
\]
Note that this is a consequence of $L_G$ and $L_B$ having the common eigenvector $\bold{1}$ with eigenvalue 0.
The unobservability of the only unstable mode implies then that the system  \eqref{eqn:systemall} is input-output stable.

\bibliographystyle{IEEETran}
\bibliography{power_sync}

\end{document}